\documentclass{PoS}

\usepackage{siunitx}

\title{Space Debris detection and tracking with the techniques of cosmic ray physics}

\ShortTitle{SD detection and tracking with CR physics techniques}

\author{\small\speaker{H.~Miyamoto}$^{1,3}$, M.~Battisti$^{1,3}$, A.~Belov$^{4}$, M.E.~Bertaina$^{1,3}$, F.~Bisconti$^1$, R.~Bonino$^{1,3}$, S.~Blin-Bondil$^{5}$, F.~Cafagna$^{6}$, G.~Cambi\`e$^{7,8}$, F.~Capel$^9$, M.~Casolino$^{7,8,10}$, A.~Cellino$^{1,2}$, I.~Churilo$^{11}$, G.~Cotto$^{1,3}$, A.~Djakonow$^{12}$, T.~Ebisuzaki$^{10}$, F.~Fausti$^{1,3}$, F.~Fenu$^{1,3}$, C.~Fornaro$^{13}$, A.~Franceschi$^{14}$,	C.~Fuglesang$^{9}$, D.~Gardiol$^2$, P.~Gorodetzky$^{15}$, F.~Kajino$^{16}$, P.~Klimov$^{4}$, L.~Marcelli$^{7}$,  W.~Marsza{\l}$^{12}$, M.~Mignone$^{1}$,  A.~Murashov$^{5}$, T.~Napolitano$^{14}$, G.~Osteria$^{17}$, M.~Panasyuk$^{4}$, E.~Parizot$^{15}$, A.~Poroshin$^{4}$, P.~Picozza$^{7,8}$, L.W.~Piotrowski$^{10}$, Z.~Plebaniak$^{12}$, G.~Pr\'ev\^ot$^{15}$, M.~Przybylak$^{12}$, E.~Reali$^{8}$, M.~Ricci$^{14}$, N.~Sakaki$^{10}$, K.~Shinozaki$^{1,3}$, G.~Suino$^{1,3}$, J.~Szabelski$^{12}$, Y.~Takizawa$^{10}$, M.~Tra\"{i}che$^{18}$,  and S.~Turriziani$^{12}$ 
for the JEM-EUSO Collaboration\\
$^1$INFN Turin, Italy; 
$^2$OATo - INAF Turin, Italy; 
$^3$University of Turin, Italy; 
$^4$SINP, Lomonosov Moscow State University, Moscow, Russia; 
$^5$Omega, Ecole Polytechnique, CNRS/IN2P3, Palaiseau, France, 
$^6$INFN Bari, Italy; 
$^7$INFN Roma Tor Vergata, Italy; 
$^8$University of Roma Tor Vergata, Italy; 
$^9$KTH Royal Institute of Technology, Stockholm, Sweden;
$^{10}$RIKEN, Wako, Japan; 
$^{11}$Russian Space Corporation Energia, Moscow, Russia; 
$^{12}$National Centre for Nuclear Research, Lodz, Poland; 
$^{13}$UTIU Rome, Italy; 
$^{14}$INFN - Laboratori Nazionali di Frascati, Italy; 
$^{15}$APC, Univ Paris Diderot, CNRS/IN2P3, CEA/Irfu, Obs de Paris, Sorbonne Paris Cit\'e, France; $^{16}$Konan University, Kobe, Japan; 
$^{17}$INFN Naples, Italy; 
$^{18}$Centre for Development of Advanced Technologies (CDTA), Algiers, Algeria \\
}

\abstract{
Space Debris (SD) consist of non-operational artificial objects orbiting around the Earth, which could possibly damage space vehicles, such as the International Space Station (ISS) or other manned spacecrafts. 
The vast majority of such objects are cm-sized, not catalogued and usually the tracking data are not precise enough. 
Here we present the feasibility study of SD detection and tracking with techniques usually employed in cosmic-ray physics. 
For this purpose, we have evaluated the possibility of using Mini-EUSO, a space-borne fluorescence telescope to be deployed on the ISS, to track SD illuminated by the Sun. 
By means of ESAF (EUSO Simulation and analysis Framework) simulation and by developing the trigger algorithms, we estimated the minimum size and maximum distances of detectable SD.
We then studied the number of possible SD detections using an ESA software called MASTER (Meteoroid and SD Terrestrial Environment Reference).
With the Mini-EUSO Engineering Model (Mini-EUSO EM), we performed some measurements to estimate the reflectance of the most common SD materials and to demonstrate the ability of Mini-EUSO to detect SD events. 
We also performed some tests in open-sky conditions, identifying and tracking fast-moving objects. In particular, the detection of a rocket body allowed us to confirm the simulation outcomes predictions and the expected performance of the detector.
}

\FullConference{36th International Cosmic Ray Conference -ICRC2019-\\
		July 24th - August 1st, 2019\\
		Madison, WI, U.S.A.}

\begin{document}\setcounter{page}{2}
\section{Introduction}
Over the last 60 years, since man began to explore space, several thousand tons of satellites and missiles have been launched and there are about 18,000 objects in orbit; ~1100 of them (6\%) are still in operation, while the remaining (94\%) can be classified as SD~\cite{ref:SD}, i.e., derelict satellites, parts of rockets and space vehicles, no longer in use, and that remain in orbit around the Earth.
These objects travel at high speeds, of the order of 7-9 km/s near the Low Earth Orbit, and can collide with spacecraft such as the ISS or other manned or unmanned spacecrafts, damaging them and in turn producing new debris.
The great majority of these objects are not catalogued and, even if they were catalogued, usually tracking data are not precise enough.
Moreover, most of them are cm-sized, that makes their detection even more difficult.

The aim of this project is to study the feasibility of SD detection and tracking with techniques usually employed in cosmic-rays physics.
We started a feasibility study investigating the performance of already existing instrumentation borrowed from the JEM-EUSO (Joint Experiment Missions for Extreme Universe Space Observatory) project~\cite{ref:JEM-EUSO}, a concept of new generation space telescopes for Extreme Energy Cosmic Ray (EECR) detection.
We benefited from the presence of the Mini-EUSO EM, a prototype of Mini-EUSO telescope~\cite{ref:Mini-EUSO}, in our lab in February and March 2018 and we performed several key tests.

SD itself do not emit the light but a Mini-EUSO-like detector can detect the reflected light from the SD illuminated by a laser or by the Sun light at sunrise and sunset (see left part of Fig. \ref{Fig:SDtriggerSchem}).
In such a way, SD can be detected as tracks crossing the Field of View (FoV) of the detector, enabling us to identify and track the SD.
This feasibility study will be also useful to verify the possibility of using an EUSO-class telescope in combination with a high energy laser for SD remediation~\cite{ref:Toshi}.
To verify this idea, we performed extended simulations and dedicated experiments at the TurLab facility located in the Physics department of the University of Turin and in open-sky conditions.

\section{Simulation}
We performed simulations to estimate the range of distances and SD dimensions detectable by a Mini-EUSO-like telescope.
The Focal Surface (FS) of such a telescope consist of a Photo-Detector-Module (PDM), which consists of 36 Hamamatsu 64-ch Multi-Anode PhotoMultiplier Tubes (MAPMTs), resulting in a readout of 2304 pixels.
For SD detection, we used a time resolution of 40.96 ms (= 1 Level 3 Gate Time Unit, 1 L3\_GTU), which corresponds to the time resolution of the Mini-EUSO level 3 (L3) data.
In this time resolution, Mini-EUSO records and store a continuous "movie" data through the entire observation time.

We simulated the light track of a SD with ESAF (EUSO Simulation and Analysis Framework), an end-to-end simulation of the phenomenon from the light emission at the source, the propagation through the environment, to the simulation of the detector response and its reconstruction algorithms (see another contribution in this conference~\cite{ref:ESAF-fenu} for the details).
We also developed the detection strategy, by testing different trigger algorithms.
The selected algorithm works offline for the moment, but could also be implemented in a Field Programmable Gate Array (FPGA) for real-time detection and active debris mitigation.
\subsection{Trigger algorithm for the SD detection}\label{sec:Trigger}
\begin{figure}
\centering
\includegraphics[width=0.94\hsize,height=6cm]{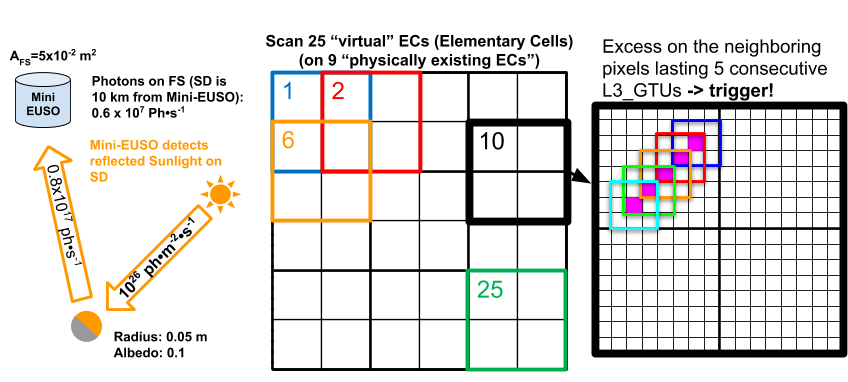}
\caption{
Conceptual figure for the SD detection by a Mini-EUSO-like detector.
The telescope detects the reflected light from SD illuminated by the Sun (Left).
Definition of virtual Elementary Cell (EC) on the PDM in the trigger algorithm (Middle). 
The trigger scans each ECs to find an excess (> $\mu_{pix}+3\times\sigma_{bkg}$) in neighboring pixels lasting 5 consecutive L3\_GTUs (Right). 
\label{Fig:SDtriggerSchem}
}
\vspace{-0.3cm}
\end{figure}

Fig.~\ref{Fig:SDtriggerSchem} shows a schematic view of the trigger logic for the SD detection by a Mini-EUSO-like detector.
Defining 25 "virtual" Elementary Cells (ECs) (middle part of Fig.~\ref{Fig:SDtriggerSchem}), the trigger scans the entire PDM and looks for an excess in neighboring pixels, which is lasting 5 consecutive L3\_GTUs (right part of Fig.~\ref{Fig:SDtriggerSchem}).
One EC consists of 4 MAPMTs. 
Neighbouring ECs are overlapping each other by 2 PMTs for vertical or horizontally, or by 1 PMT for diagonally.
With a threshold of pixel count, which is 3 $\sigma$ above the average background in the pixel, $\mu_{pix}+3\times\sigma_{bkg}$, the fake trigger rate becomes low enough as $<3\times10^{-6}$ Hz.
\subsection{Maximum Detection Distance}
Using ESAF simulation and applying externally the trigger algorithm described above to the simulated data, we estimated the maximum distance that a Mini-EUSO-like telescope can detect SD.
The plot in the Fig.~\ref{Fig:maxDistanceSat} at the end of this paper shows the estimated maximum detection distance [km] as a function of the size of SD in radius [cm]. 
Here the value 0.5 for the SD reflectance is employed in the simulation, which is confirmed to be more or less appropriate by the lab and open-sky tests as described in later sections.
\subsection{Estimation of the expected number of SD detection by Mini-EUSO} 
After estimating the maximum distance for SD detection as a function of SD size, we estimated the expected number of SD detection by a Mini-EUSO-like telescope using MASTER~\cite{ref:MASTER}, a simulation software developed by ESA which provides us with the density of SD with parameters such as altitude and inclination.
We integrated the total number of SD provided by MASTER within the telescope FoV ($\pm18^{\circ}$ at the orbital inclination of $\pm51.64^{\circ}$ with an altitude of $\sim$400km, and the depth of maximum distances described above) multiplied by the observation time.
Here we employed 5 min per orbit for the observation time, which corresponds to the twilight time. 
For the smallest size of SD (r=0.5[cm]) simulated here a time resolution of 320 $\mu$s is used instead of 40.96 ms, which corresponds to the one for Mini-EUSO Level 2 data, as it can be detected only when it is relatively closer to the telescope.

Table~\ref{tb:nDet} shows the result of the calculation.
\begin{table}
\centering
\begin{tabular}{|c|c|c|c|c|}
\hline
\begin{tabular}{c}
SD radius\\
\[[cm]\\
\end{tabular}
& 
\begin{tabular}{c}
altitude\\
\[[km]\\
\end{tabular}
& 
\begin{tabular}{c}
distance from\\
Mini-EUSO\\
\[[km]\\
\end{tabular}
& 
\begin{tabular}{c}
ave. N$_{SD}[/km^3]$ for\\  
declination $\pm51.64[^\circ]$,\\
refl=0.5\\
\end{tabular}
& 
\begin{tabular}{c}
\\
N$_{det}/day$\\
\\
\end{tabular}\\
\hline
5.0 & 300 & 100 & 1.91E-09 & 1.47E-01\\
\hline
4.5 & 312 &   88 & 5.68E-13 & 3.61E-05\\
\hline
4.0 & 323 &   77 & 6.33E-13 & 3.61E-05\\
\hline
3.5 & 334 &   66 & 7.17E-13 & 2.56E-05\\
\hline
3.0 & 344 &   56 & 3.28E-11 & 8.42E-04\\ 
\hline
2.5 & 352 &   48 & 5.02E-10 & 9.47E-03\\ 
\hline
2.0 & 365 &   35 & 8.93E-10 & 8.96E-03\\ 
\hline
1.5 & 373 &   27 & 1.41E-08 & 8.42E-02\\ 
\hline
1.0 & 385 &   15 & 1.53E-08 & 2.81E-03\\ 
\hline 
0.5 & 396 &     4 & 3.42E-07 & 4.48E-02\\ 
\hline 
	 &	  &	    & TOTAL(N$_{det}$/day)& 3.08E-01\\
\hline
	 &	  &	    & TOTAL(N$_{det}$/yr)   & 112\\
\hline
\end{tabular}
\caption{Maximum detection distance and expected number of the SD detection by Mini-EUSO, estimated with .
For the minimum size of SD (r=0.5cm), the time resolution of Mini-EUSO level 2 data (=320 $\mu$s) is used instead of the one for level 3 (=40.96 ms).
\label{tb:nDet}
}
\vspace{-0.5cm}
\end{table}
The obtained value above is not taking into account the effect of increasing background level due to the twilight.
During the night-sky observation described in section~\ref{Sec:NightSkyObservation}, the background level is increased by a factor of 3.
This fact directly affects the maximum detection distance, lowering it by a factor of 1/$\surd{3}$.
Therefore, the final value for the expected number of SD detection taking into account this effect would be $\sim$45 debris/yr. 
\subsection{Tracking SD}
\begin{figure*}
\vspace{-0.5cm}
\centering
\includegraphics[width=0.97\hsize]{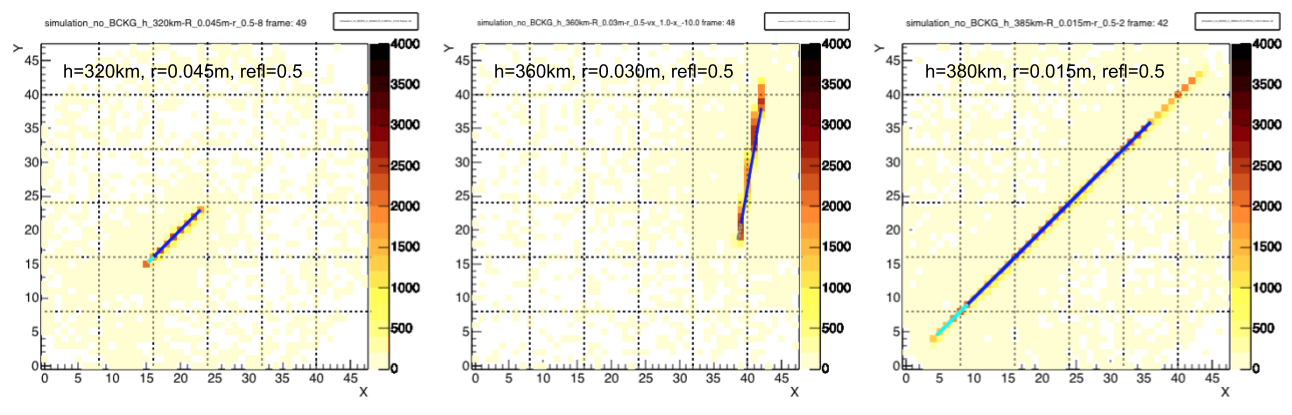}
\caption{Examples of SD tracking algorithm applied to ESAF simulation data of SD with a reflectance of 0.5, different radii (4.5, 3.0, 3.5 [cm]) and altitudes (320, 360, 380 [km]).
Blue line shows the track derived from the first and the last triggered point, light blue line shows the predicted track for 5 GTUs after the last triggered point.
\label{Fig:SDtracking}
}
\vspace{-0.5cm}
\end{figure*}

After obtaining triggered events, we tried to implement the SD tracking using the trigger output data, and to predict the SD position some time later.
The trigger output file provides us the position of SD on the detector FS, together with the time stamp in a unit of L3\_GTU (=40.96 ms).
Then we derive the track of SD as a linear function of a time (L3\_GTU).
Fig.~\ref{Fig:SDtracking} shows examples of the tracking and prediction method applied to the data simulated by ESAF.
The dark blue line along the track is the calculated track from the trigger output file, while the continued light blue line shows the predicted track in the following L3\_GTUs.
We tried several kinds of events with different crossing angles and altitudes of SD.

The study of the estimation of the maximum and minimum distance of an SD, as well as the estimation of the minimum number of L3\_GTUs required to predict the track sufficiently well, is currently ongoing.
%
%
\section{Mini-EUSO EM tests at TurLab and open-sky conditions}
We performed some laboratory measurements at the TurLab facility~\cite{{ref:TurLab}} as well as in an open-sky condition at the observatory in Pino-Torinese (see other contributions to this conference~\cite{ref:Bisconti},~\cite{ref:EUSOatTurLab} for the details).  
%
We prepared the experimental setup at the TurLab, that is equipped with a rotating tank ($\phi$ = 5 m, h = 1 m) and represents an ideal condition for testing SD detection, being much darker than the night sky by several orders of magnitude.

For these tests, we used the rotating tank with a series of configurations to reproduce the Earth views.
Mini-EUSO EM was hung from the ceiling above the tank pointing downward to mimic the detector's view from the ISS.
At that time, the Mini-EUSO EM had only 4 central MAPMTs and we used a 1" $\phi$ plano-convex lens instead of the Fresnel lenses for all the measurements described in this paper.
In this configuration, the telescope has a FoV of 0.5$^{\circ}$/pix instead of 0.8$^{\circ}$/pix for the Mini-EUSO telescope.
Two strips of cold-white LEDs illuminating the ceiling were used to reproduce the diffused light above those materials.
In this way, we could perform the measurements in a controlled way, for different levels of the background and reflected light from the materials, depending on the detector configuration or the conditions we wanted to test. 
\subsection{SD measurement with Mini-EUSO EM at TurLab}
\begin{figure*}
\centering
\vspace{-0.5cm}
\includegraphics[height=6.8cm]{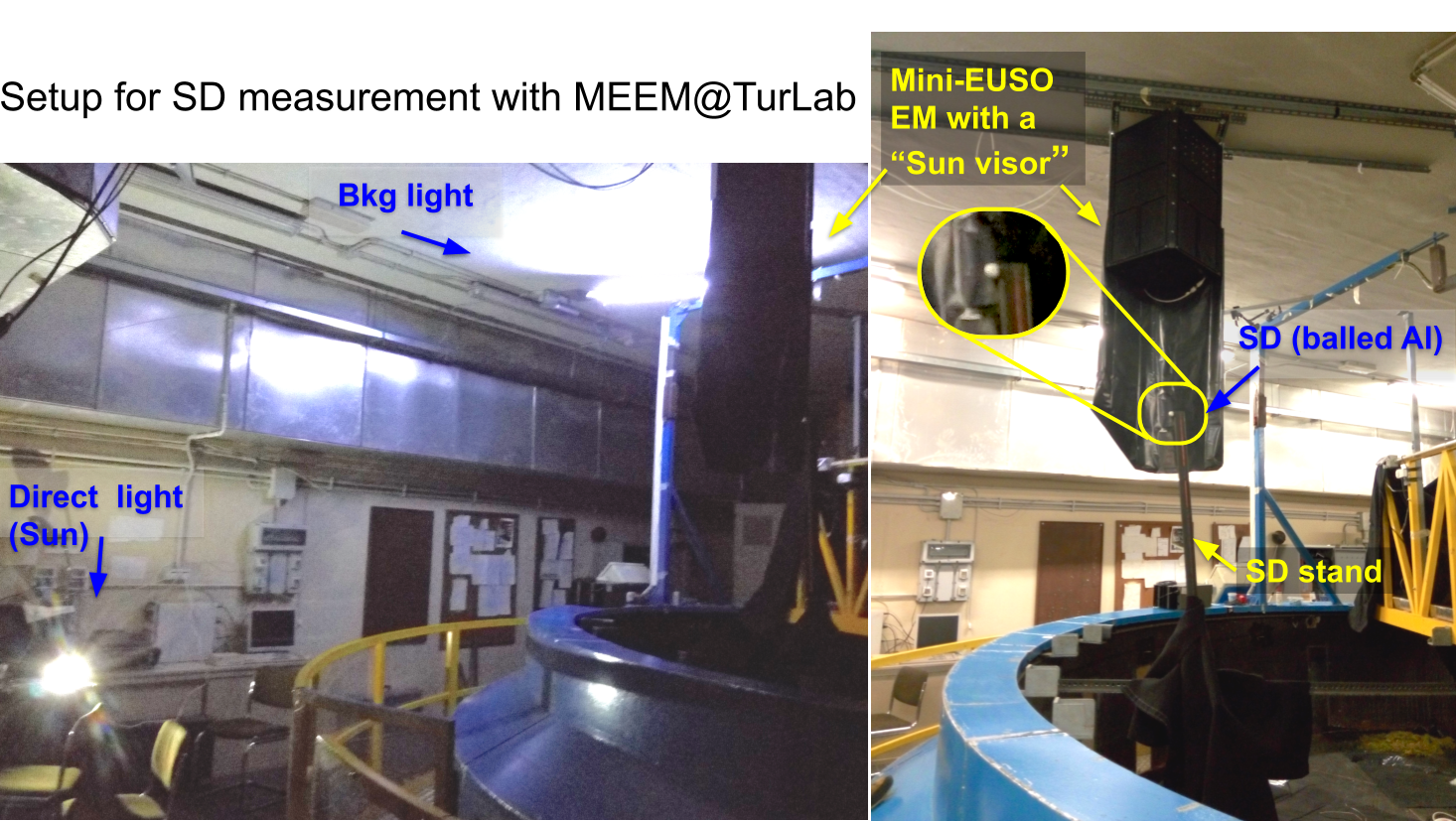}
\caption{TurLab setup for SD detection by the Mini-EUSO EM. The detector FS is protected by a "Sun visor" to avoid the direct light from the white LED which is fixed outside of the tank, mimicking the Sun light. A balled Aluminium under the telescope is attached to a stand which is fixed to the bottom of the tank. As the tank rotates, the balled Aluminium moves within the detector FoV being illuminated by the LED.
\label{Fig:SDatTurLab}
}
\end{figure*}

\begin{figure*}
\vspace{-0.3cm}
\centering
\includegraphics[height=6.8cm]{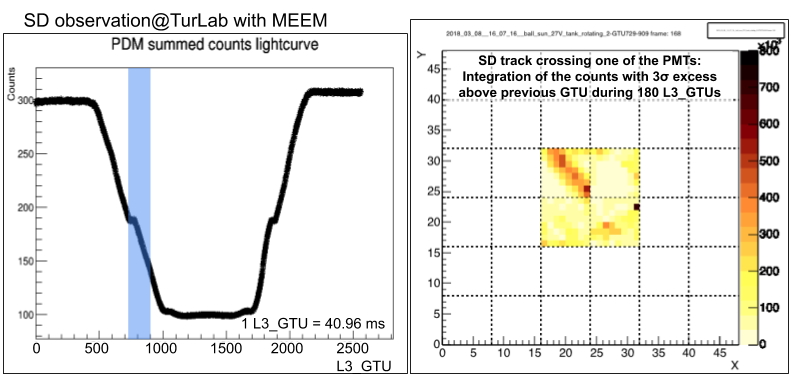}
\caption{
Left: Summed L3 data of all 4 PMTs as a function of L3\_GTU during the balled Aluminium passing back and forth the Mini-EUSO EM FoV.
Right: SD track crossing one of the PMTs. L3 data with an excess of 3 $\sigma$ above the counts in previous L3\_GTU are integrated during 180 L3\_GTUs (blue coloured area in the left plot).
\label{Fig:SDevent}
}
\vspace{-0.5cm}
\end{figure*}

Fig.~\ref{Fig:SDatTurLab} shows the setup for reproducing SD detection principle.
Mini-EUSO EM is hung on the ceiling above the TurLab tank, with a "Sun visor" to avoid the direct light from the high power LED which is mimicking the Sun light.
A balled Aluminium under the telescope is attached to the edge of a stand which is fixed to the bottom of the tank.
As the tank rotates, the balled Aluminium moves within the detector FoV being illuminated by the LED,
while the Mini-EUSO EM FS remains in the shade of the Sun visor.
The right plot of Fig.~\ref{Fig:SDevent} shows the image of the track of the balled Aluminium moving within the detector FoV, during the time (in units of L3\_GTU=40.96 ms) highlighted in the light curve on the left. The image is integrated in a way similar to our trigger to obtain the track, i.e., keeping the counts in each pixel when it is exceeding 3$\sigma$ above the one in the previous L3\_GTU.
\subsection{SD reflectivity measurement}
Different materials such as polished Aluminium foil, canned Aluminium, unpolished Aluminium foil, mirror, copper foil, white paper, balled Aluminium foil, a sample of Kevlar$^{\footnotesize{\textregistered}}$, electronics board, often make the SD, were placed on the bottom of the tank, within the Mini-EUSO EM FoV, in order to measure the relative reflectances of these materials as seen by the detector.
For the rough estimation, assuming that the reflectivity of polished Aluminium is about 0.92~\cite{ref:PrecisionOptics}, 
we observed the reflectance of the different materials varying from 0.17 (electronics board) to 0.92 (fixed reference value for the polished Aluminium foil).
The mean value $\sim$0.5 was used for the ESAF simulation.
It also seems consistent with a rocket body detection during the open-sky measurement as described in the following section.
\subsection{Night-sky observation and a rocket body detection}\label{Sec:NightSkyObservation}
\begin{figure*}
\vspace{-0.2cm}
\centering
\includegraphics[width=0.9\hsize,height=8cm]{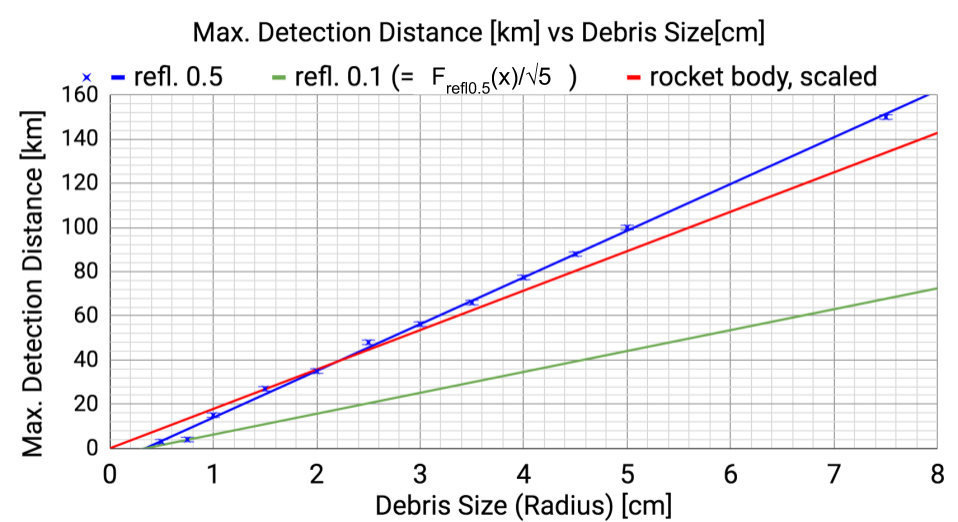}
\caption{
Blue line with data points: Maximum detection distance as a function of debris size in radius, 
obtained by ESAF simulation applying externally the trigger algorithm described in section\protect\ref{sec:Trigger}.
Red line: Scaled down distance as a function of size of the detected rocket body.
Green line: Scaled red line down to the case of the SD reflectance of 0.1.
}\label{Fig:maxDistanceSat}
\vspace{-0.3cm}
\end{figure*}

We also performed night-sky observations at the Astronomy Observatory in Pino-Torinese, using Mini-EUSO EM in March 2018.
A rocket body, as well as several stars, has been detected during the observation period.
Scaling down the size and distance of the detected rocket body, we could estimate the actual performance of Mini-EUSO EM against SD-like objects.
With such a scaling, we could see if the simulation is reasonable or not.
The detected object turned out to be Meteor 1-31 Rocket, a cylindrical object 2.8 m $\times$ 2.6 m on a $\sim$530 km orbit.
The red line in the Fig.~\ref{Fig:maxDistanceSat} shows the altitude (distance from the telescope) and size of the detected rocket body scaled to an object in the range of the plot, considering the lens size, pixel FoV and the absorption by the air.
For example, the detected photon counts of this rocket body with its size and altitude corresponds to an object with the size of 5.6 cm if it were at the distance of 100 km, or to the size of 2.8 cm if it were at a distance of 50 km from the telescope. 

Also, the case for the reflectance of 0.1 is estimated by scaling the result of 0.5 as shown as the green line in the Fig.~\ref{Fig:maxDistanceSat}, as our detection strategy purely depends on the luminosity which is proportional to the area of SD.
One may see that the reflectance could be a bit smaller than 0.5 but certainly much higher than 0.1.
These results imply that our simulation and estimation of the reflectance are conceivable with the actual situation (see Fig. ~\ref{Fig:maxDistanceSat}).
\section{Summary}
We performed the feasibility study with simulations, laboratory and open-air measurements of the prototype Mini-EUSO telescope.
The ESAF and MASTER simulations applying with our trigger algorithm indicated that we are capable of observing SD, and for the case of Mini-EUSO, we estimate that we will detect $\sim$45 SD per year.

The SD tracking and prediction algorithm is also studied using the ESAF simulation and its trigger output file.
We found that as a first step, the estimated track and prediction are well matching to the SD tracks.
Further studies in different conditions, limit of the tracking and prediction in the same logic are currently ongoing.
We also performed reflectance measurements of various SD materials, that provides us with the mean value of about 0.5.

We detected a rocket body during a night-sky observation with Mini-EUSO EM.
Conversion of the detected object to the size and distance, makes it comparable to SD with the size of 5.6 cm in radius, at the distance of 100 km, or to the SD with the size of 2.8 cm in radius, at the distance of 50 km from the telescope.
Such values are well along the result of maximum SD detection distance in the case of SD reflectance  is 0.5 which implies that at least the simulations are predicting detections in the right range of sizes and distances.
\vspace{-0.2cm}
\section*{Acknowledgments}
\vspace{-0.2cm}
\noindent The support received by the Astronomical Observatory of Turin is deeply acknowledged.\\
The authors acknowledge support from Compagnia di San Paolo with the project ``New techniques for the detection of space debris''; Id Project: CSTO164394.\\
This work was partially supported by Basic Science Interdisciplinary Research Projects of RIKEN and JSPS KAKENHI Grant (22340063, 23340081, and 24244042), by the Italian Ministry of Foreign Affairs and International Cooperation, by the Italian Space Agency through the ASI INFN agreement n. 2017-8-H.0, by NASA award 11-APRA-0058 in the USA, by the French space agency CNES, by the Deutsches Zentrum f\"{u}r Luft- und Raumfahrt, the Helmholtz Alliance for Astroparticle Physics funded by the Initiative and Networking Fund of the Helmholtz Association (Germany), by Slovak Academy of Sciences MVTS JEM-EUSO, by National Science Centre in Poland grant (2015/19/N/ST9/03708), by Mexican funding agencies PAPIIT-UNAM, CONACyT and the Mexican Space Agency (AEM), as well as VEGA grant agency project 2/0132/17, and by State Space Corporation ROSCOSMOS and Russian Foundation for Basic Research (grant 16-29-13065).

\end{document}